\documentclass[11pt,twoside]{article}

\usepackage{asp2006}
\usepackage{epsf}
\usepackage{graphicx}
\usepackage{lscape}

\begin{document}
\title{The Evolution of Stellar Mass Density and its Implied
Star Formation History}
\author{S. M. Wilkins, N. Trentham}
\affil{University of Cambridge}
\author{A. M. Hopkins}
\affil{University of Sydney}

\begin{abstract}
Using a compilation of measurements of the stellar mass density
as a function of redshift we can infer the cosmic star formation
history. For $z < 0.7$ there is good agreement between the two star
formation histories. At higher redshifts the instantaneous indicators
suggest star formation rates larger than that implied by the evolution
of the stellar mass density. This discrepancy peaks at $z = 3$ where
instantaneous indicators suggest a star formation rate around 0.6 dex
higher than those of the best fit to the stellar mass history. We discuss
a variety of explanations for this inconsistency, such as inaccurate
dust extinction corrections, incorrect measurements of stellar masses
and a possible evolution of the stellar initial mass function.
\end{abstract}
\vspace{-1cm}

\section{Introduction}
\vspace{-2mm}
Much contemporary research in extragalactic astronomy has revolved around
the determination of the instantaneous cosmic star formation history
\cite[SFH,][]{Lil:96,Mad:96}. Measuring this quantity from observations
requires a number of assumptions, with the form of the dust obscuration
corrections and stellar initial mass function \cite[IMF, e.g.,][]{Kro:07}
being among the most important. Integration of the instantaneous SFH over
redshift, making appropriate corrections for stellar evolution processes,
yields the current stellar mass density. This quantity can be independently
measured, and numerous studies have attempted comparisons of
these quantities \cite[e.g.,][]{MPD:98,Col:01,Fon:04,Arn:07}.
A number of studies \cite[HB06 hereafter]{Eke:05,HB:06} also claim the
instantaneous SFH overpredicts the low-redshift stellar mass density.
We investigate this possible discrepancy using a compilation
of the most up to date measurements of the stellar mass-density history (SMH).

A compilation of both low- and high-redshift measurements of the stellar
mass density are given by \cite{WTH:08}. Using these values we
derive a best-fitting SFH, comparing this estimate to other estimates of
the SFH and highlighting the discrepancies. We assume
$H_0=70\,$km\,s$^{-1}$\,Mpc$^{-1}$, $\Omega_M=0.3$, $\Omega_{\Lambda}=0.7$.
\vspace{-6mm}

\section{Comparing the SMH with the SFH}
\begin{figure}[t]
\centerline{\rotatebox{0}{\includegraphics[width=10.0cm]{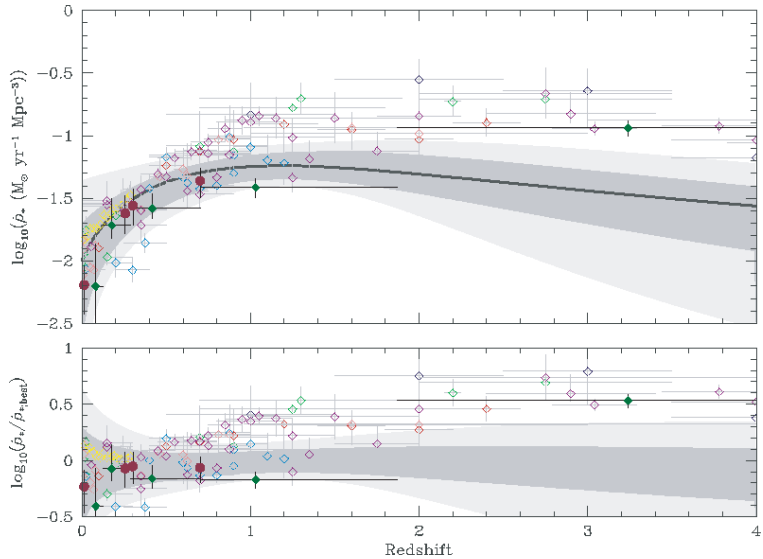}}}
\caption{Comparison of the SFH inferred from the SMH compared to direct
measurements. The SFH inferred from the evolution of stellar mass
is shown by the $1\,\sigma$ and $3\,\sigma$
uncertainty regions (dark and light grey-shaded areas, respectively). The dark
solid line is the parameterised best fit derived from the SMH. The
lower panel displays the residuals from this best-fitting SFH of the other
measurements. See \cite{WTH:08} for details.
 \label{fig:sfh}}
\vspace{-5mm}
\end{figure}

\vspace{-2mm}
The SMH, $\rho_*(t)$, can be expressed as the integral of the SFH,
$\dot{\rho}_*(t)$, corrected for the effects of mass-loss through stellar
evolution processes such as supernovae and stellar winds. This relation
may be inverted to determine the SFH from the observed evolution of stellar
mass, as described by \cite{WTH:08}. Fig.\,\ref{fig:sfh} shows a best-fitting
SFH derived from the SMH compilation and using the formalism of \cite{WTH:08},
along with $1\,\sigma$ and $3\,\sigma$ uncertainty regions.
For $z < 0.7$ the $1\,\sigma$ uncertainty region of this SFH is consistent
with the best-fitting instantaneous SFH obtained by HB06. For $z > 0.7$ the
best-fitting SFH of HB06 is consistently higher than that inferred from
the SMH. At $z = 3$
the best fit of HB06 SFH implies a star formation rate around four
times (0.6 dex) larger than that inferred from the stellar mass density.

\begin{figure}[t]
\centerline{\rotatebox{0}{\includegraphics[width=8.6cm]{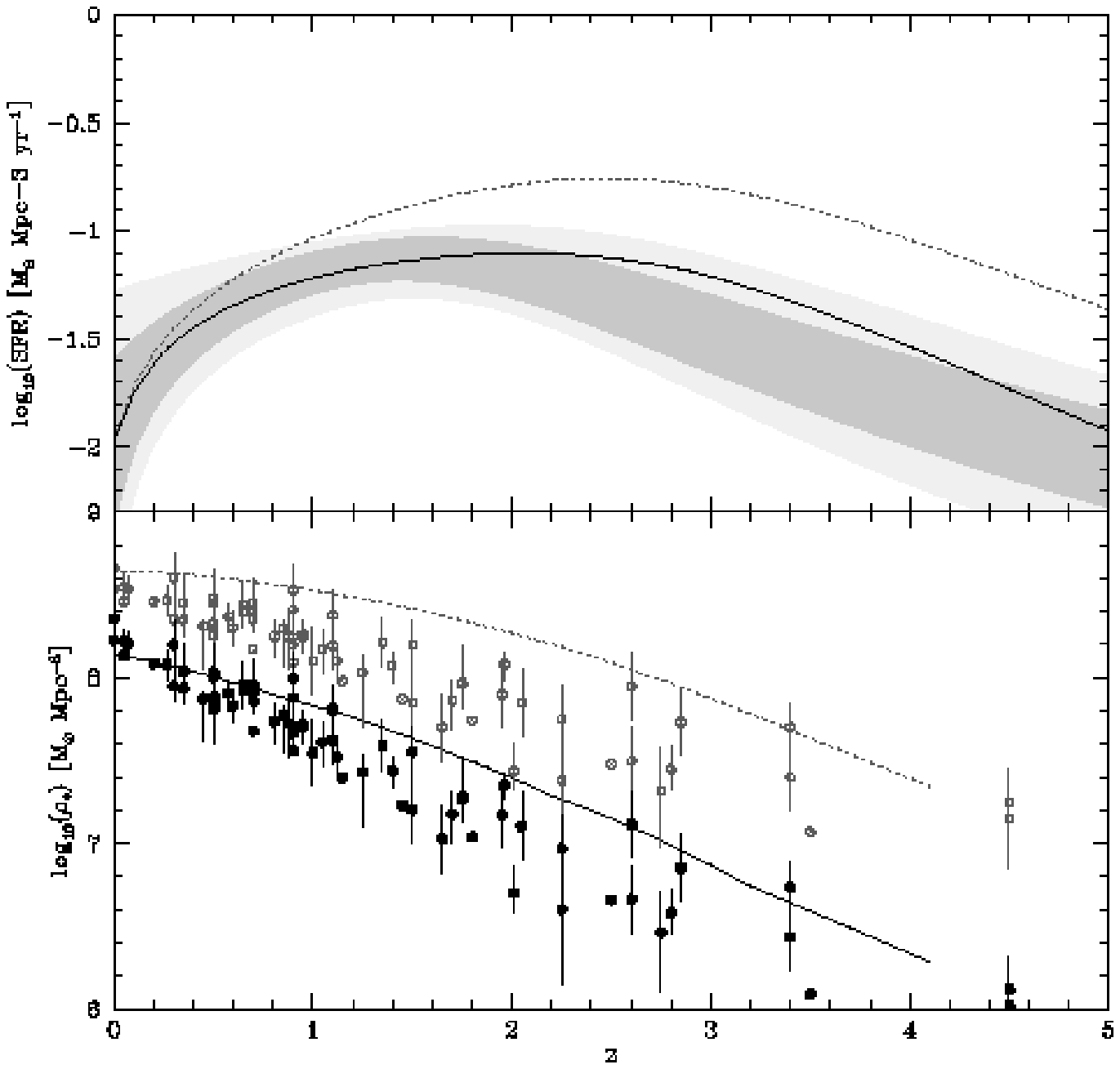}}}
\caption{Top: The best-fitting instantaneous SFH to the compilation
of HB06 obtained with a universal IMF (dotted line) compared with that
obtained with an evolving IMF (solid line). Shaded regions are
the $1\,\sigma$ and $3\,\sigma$ uncertainty regions.
Bottom: The observed SMH assuming a universal IMF (open circles) and
assuming an evolving IMF (filled circles) and corresponding predictions
from the instantaneous SFH.
 \label{fig:smh}}
\vspace{-5mm}
\end{figure}

This discrepancy suggests that either stellar mass estimates are incorrect
or the SFH of HB06 is overestimated at $z > 0.7$, or perhaps both.
These possibilities are explored in some detail by \cite{WTH:08}.
An alternative solution is that the larger star formation rates could
be explained by an evolution of the star formation rate calibration,
such as that from an evolving IMF. Investigation of an environmental
or temporal evolution of the IMF has been carried out by a number
of authors \cite[e.g.,][]{Nag:05,Bau:05,LeD:06,Lac:07}.
\cite{Far:07} also investigates the possibility of a
different IMF in starburst galaxies to explain a discrepancy between
the extragalactic background light, the instantaneous SFH and the
K-band luminosity density. An IMF that produces more emission associated
with instantaneous indicators (such as the UV or H$\alpha$ luminosity)
per unit mass created is required at high redshift. This implies evolution
toward a more ``top-heavy" (high-mass biased) IMF with increasing redshift.
\cite{WTH:08} introduce such an evolving IMF. This evolution in the IMF
changes the SFH implied by instantaneous indicators, the fraction of
material recycled as a function of age and the observed stellar mass density.
This simple model provides increased agreement between
the observed SMH and that inferred from the SFH, shown in Fig.\,\ref{fig:smh}.

Many other forms of evolving IMF may also reproduce the relationship
between the instantaneous SFH and the SFH derived from the evolution
of stellar mass. The model shown here simply illustrates the effect of
an increasing high-mass bias in an IMF toward high redshift, showing that such
an effect is sufficient to explain the discrepancies between the
instantaneous SFH and the SFH inferred from the SMH. It is likely that
real stellar IMFs may have more complex evolution, and this is
being investigated in ongoing work.
\vspace{-5mm}

\section{Summary}
\vspace{-2mm}
A compilation of stellar mass density measurements spanning
$0 < z < 4$ was compared to the SFH, by deriving a
best-fitting SFH, well described by the \cite{Col:01}
parameterisation with $a = 0.014$, $b = 0.11$, $c = 1.4$ and $d = 2.2$.
There is good agreement at $z<0.7$ between these estimates, but at
progressively higher redshifts the stellar mass density and instantaneous
indicator inferred SFHs become inconsistent. Instantaneous measures at
$z = 3$ imply best-fitting star formation rates four times larger than those
inferred from the SMH.
There are a number of possible causes of this tension. These include
possible systematic effects in stellar mass-density estimation, and
uncertainty in the effects of dust on both stellar mass
estimates and high-redshift star formation rate estimates.

A more speculative solution is an effective temporal evolution of the IMF.
We have identified a simple, non-unique
model for an evolving IMF that reconciles both the SFH and SMH.
Other recent evidence for an evolving IMF has been explored by
\cite{Dav:08} and \cite{vD:08}, who provide different parameterisations.
Given the importance of both stellar mass density and star formation rate
density measurements to our understanding of the galaxy formation process
it is crucial that this discrepancy be resolved. To achieve this,
refinements need to be made to measurements, and extensions, such
as an evolving IMF, to galaxy formation models need to be implemented.
Measurements of the SMH can be improved through larger and deeper surveys,
more thorough understanding of dust attenuation, and improvements
in population synthesis models. Improvements in the SFH can also be
made by better observational constraints on dust attenuation,
and independent estimates such as those from supernova rate densities,
fossil-histories from local galaxies, gamma-ray bursts,
and the diffuse supernova neutrino background.
\vspace{-2mm}

\acknowledgements
SMW acknowledges support of an STFC studentship and of King's College,
NT acknowledges support provided by STFC and AMH acknowledges support provided
by the Australian Research Council in the form of a QEII Fellowship
(DP0557850).\\
\vspace{-3mm}

\noindent {\bf Comments}\\
{\it A.\ Renzini}: A flat IMF at $z\approx 3$ would overproduce metals in
galaxy clusters by a sizeable factor.\\
{\it D.\ Wilman}: It would be interesting to consider a varying IMF as a
function of a physical parameter, such as the SFR. Pavel Kroupa has been
suggesting recently that this might be the case.\\
{\it M.\ Dickinson}: Something physical about galaxies that affects their
IMF is a more preferable driver than simply redshift. It would be useful
to predict other observable, testable properties that could be used to
test any given prediction for an evolving or varying IMF.

\end{document}